\documentclass[12pt,preprint]{aastex}

\newcommand{\myeq}[1]{Eq. \ref{#1}}
\newcommand{\myfig}[1]{Fig. \ref{#1}}

\begin{document}

\title{Optimal Occulter Design for Finding Extrasolar Planets}

\author{Robert J. Vanderbei}
\affil{Operations Research and Financial Engineering, Princeton
University}
\email{rvdb@princeton.edu}
\author{Eric Cady}
\affil{Mechanical and Aerospace Engineering, Princeton University}
\email{ecady@princeton.edu}
\and
\author{N. Jeremy Kasdin}
\affil{Mechanical and Aerospace Engineering, Princeton University}
\email{jkasdin@princeton.edu}

\begin{abstract}
One proposed method for finding terrestrial planets
around nearby stars is to use two spacecraft---a telescope and a specially
shaped occulter that is specifically designed to prevent all but a 
tiny fraction of the starlight from diffracting into the telescope. 
As the cost and observing cadence for such a mission will be driven largely by
the separation between the two spacecraft, it is critically important to
design an occulter that can meet the observing goals while flying as close to
the telescope as possible.  In this paper, we explore this tradeoff between
separation and occulter diameter.  More specifically, 
we present a method for designing the shape
of the outer edge of an occulter that is as small as possible and gives 
a shadow that is deep enough and large enough for a $4$m telescope 
to survey the habitable zones of many stars for Earth-like planets.
In particular, we show that in order for a $4$m telescope to detect in 
broadband visible light a planet $0.06$ arcseconds from a star 
shining $10^{10}$ times brighter 
than the planet requires a specially-shaped occulter $50$m 
in diameter positioned about $72,000$ km in front of the telescope.
\end{abstract}


\section{Introduction} \label{sec:intro}

Since the first extrasolar planet was discovered by
\cite{May95}, various techniques have been used to
infer the presence of large planets, but none have the capability to
image Earth-like planets directly. Finding terrestrial planets is
difficult because the difference in brightness between the star and
the planet is so large; in the visible spectrum, the intensity
difference, or contrast, is approximately $10^{10}$ (see \cite{Des02}).
Just as importantly, the maximal angular separation between planet and star
is on the order of $0.1$ arcseconds for a star 10 parsecs from
Earth. The problem becomes one of reducing the intensity of
starlight at the planet's location by a factor of $10^{10}$. Several
methods (see \cite{Kas03,Kuc02,Van06,Guy03})
have been proposed to do this within the telescope
by adjusting the point spread function so that there is very little
starlight at the location of the planet in the image plane of the
telescope.  While these techniques have demonstrated the potential
to provide the necessary contrast (\cite{Sid06}), they have the
intrinsic difficulty that they require an adaptive-optics system
within the telescope to correct aberrations in the wavefront (induced largely
by imperfect optics), which
tend to spill unwanted light into the search area.

One solution to this problem is to remove the starlight before it
reaches the telescope by using a second spacecraft, an
\emph{occulter}, positioned between the telescope and the target
star. Such a concept was first proposed by
\citet{Spi62}; since then, a number of proposals
(see \cite{Cop00,Sch03,Cas06}) have appeared
that use occulters to look for planets, both of Jupiter and
Earth size.

Simple ray optics would suggest that a circular disk occulter of
diameter $D$ would be adequate to block all of the starlight from entering a
telescope of aperture $D$.
Unfortunately, this analysis
neglects diffraction, which is a significant factor in propagations
involving narrow angles. It was known as early as 1818 that diffraction
around objects could produce light in areas that geometric optics
would predict to be dark; the most notable example of this is Poisson's
spot, which earned Fresnel a prize from the French Academy of
Sciences (see \cite{Goo96}).  A proper design of an occulter-based
mission thus requires careful consideration of diffraction
effects.

\citet{Spi62} noted that it was sufficient to change the
transmission function with radius in order to suppress this central
spot.  Subsequent papers have suggested some specific mechanisms for
accomplishing this.  One concept, called the {\em Big Occulting Steerable
Satellite (BOSS)} (\cite{Cop00}), 
is based on a transmissive apodization defined by polynomials.
A more recent entry into this field is the {\em New Worlds Observer} proposed
by Webster Cash and funded by NIAC.  Originally conceived as a pinhole camera
in space (\cite{2004SPIE.5487.1634S}), it was eventually reincarnated as a
space-based occulter (\cite{Sim04,Cas06}).
There is some hope that an occulter mission, if technically feasible, could
overcome the significant challenges that more traditional coronagraphic
approaches to planet finding must confront.
The purpose of this paper is to
explore the trade-off between inner-working-angle 
and telescope-occulter separation.  We show that an occulter capable of
detecting in broad-band visible light 
an Earth-like planet at 60 mas separation from its parent star will
need to be 50m in diameter (tip-to-tip) and fly 72,000 km in front of the
telescope.
%

\section{Babinet's principle} \label{sec:prin}

An occulter is complementary to a pinhole
camera; instead of allowing light only through a small hole, an
occulter allows all light except for the light blocked by the occulter which
now replaces the small hole. This
complementarity allows us to calculate the downstream electric field produced
by an occulter using Babinet's principle; that is, the sum of the light
passing around the occulter and the light passing through an
occulter-shaped hole is a free-space plane wave. The electric field
past the occulter is thus given by
\begin{equation}\label{oc1}
    E_{o} = E_{u} -E_{h}
\end{equation}
where $E_o$ is the field produced by an occulter, 
$E_u$ is the electric field of an unobstructed plane wave, and 
$E_h$ is the field produced by a complimentary pinhole. 
From the Helmholtz equation it follows that a plane wave having 
complex amplitude $E_0$ at the plane of the occulter would, if unimpeded by an
occulter, be given by
$E(\rho,\phi) = E_0 e^{2 \pi i z / \lambda}$ at the telescope's pupil plane,
which is located a distance $z$ behind the occulter.  Here, and
throughout the paper, we use polar coordinates $(\rho, \phi)$ to represent the
pupil plane of the telescope.  We assume that $\rho = 0$ corresponds to the
center of the pupil.

Before we investigate simple shaped occulters, it is instructive to consider a
more general setting in which an occulter (or a hole) need not be purely opaque
or transparent.  Instead, we introduce the possibility for partial
attenuation.  To this end, we introduce a function $A(r,\theta)$ to denote the
attenuation profile for the occulter (we use $r$ and $\theta$ to denote polar
coordinates in the plane of the occulter).  If $A(r,\theta)$ takes the value $1$
somewhere, then no light gets through at that point.  On the other hand, if
it takes the value zero, then all light gets through.  All values in between
are allowed.
Under circular symmetry, the attenuation profile $A(r,\theta)$ does not depend
on $\theta$ and so we can write
$A(r)$ for the attenuation profile.
Of course, when thinking about transmission through a ``tinted'' hole, the
function $A(r)$ represents the level of transmission rather than attenuation.
That is, $A(r)$ describes an apodization of the hole.
With these assumptions, $E_{h}$ at the occulter plane can be written as
\begin{equation} \label{oc13}
    E_{h}(r) = E_{u}A(r) = E_0A(r) .
\end{equation}
Assuming that the function $A(r)$ is zero for $r$ larger than some threshold
$R$, the Fresnel integral for the propagation of the field from the hole
a distance $z$  can then be written in polar coordinates
(\cite{Goo96}) as
\begin{equation}\label{oc15}
    E_{h}(\rho) 
    = 
    E_0
    \frac{2 \pi }{i \lambda z}
    e^{\frac{2 \pi i z}{\lambda}}
    e^{\frac{\pi i \rho^2}{\lambda z}}
    \int^R_0 
        J_0\left(\frac{2 \pi r \rho}{\lambda z}\right)A(r)
	e^{\frac{\pi i}{\lambda z}r^2}
        r dr
\end{equation}
and the field due to an occulter can be expressed as
\begin{equation}\label{oc16}
    E_{o}(\rho) 
    = 
    E_0e^{\frac{2 \pi i z}{\lambda}}
    \left(
        1
	-
	\frac{2 \pi e^{\frac{\pi i \rho^2}{\lambda z}}}{i \lambda z}
        \int^R_0 
	    J_0\left(\frac{2 \pi r \rho}{\lambda z}\right)A(r)
	    e^{\frac{\pi i}{\lambda z}r^2} 
	    r dr
    \right) .
\end{equation}

\section{Optimal attenuation functions} \label{sec:optim}

We find $A(r)$ by minimizing the ``extent'' of $A(r)$ subject to the
constraint that the intensity, which is the square of the magnitude of the
electric field, in a specified dark region is no more
than $10^{-10}$.  To be precise, we 
%
%
%
%
%
\begin{eqnarray}
    \mbox{minimize } & \int^R_0 A(r) r dr & \nonumber \\
    \mbox{subject to } 
      & |E_o(\rho)|^2 \leq 10^{-10}|E_0|^2, 
        & \quad 0 \leq \rho \leq \rho_{\mathrm{max}} \nonumber \\
      & 0 \leq A(r) \leq 1, & \quad 0 \leq r \leq R. 
    \label{op4}
\end{eqnarray}
This is an infinite-dimensional, quadratic programming problem,
which would produce a shadow from $0$ to $\rho_{\mathrm{max}}$ with
$10^{10}$ contrast at the telescope's pupil plane. Unfortunately, it is
computationally intractable.  To make it solvable, we introduce
certain simplifications to reduce it to a finite-dimensional, linear
programming problem.

First, we rewrite the constraint on $E_o(\rho)$ as:
\begin{equation} \label{op5}
    |E_o(\rho)| \leq 10^{-5}|E_0|
\end{equation}
Since $E_o(\rho)$ is complex, we can constrain
the magnitude of the real and imaginary parts of $E_o(\rho)$ to be
less than or equal to $10^{-5} |E_0|/\sqrt{2}$ to get a more
conservative, but linear, constraint on $E_o(\rho)$.  Finally, we
discretize $[0, R]$ and $[0, \rho_{\mathrm{max}}]$ to get a
finite-dimensional program.

As formulated, this optimization model produces the desired shadow only
at a single selected wavelength.  For such an optimization model,
the ``optimal'' function $A(r)$ turns out to take on only two values: zero and
one.
In other words, the solution is a {\em concentric ring mask} (see
\cite{VSK02}).  Such a solution
achieves the desired contrast at the specified wavelength, but its
performance degrades quickly as one moves either to longer or
shorter wavelengths.  To find a design that works over a broad band of
wavelengths, we make a few simple
changes to our optimization model.  Obviously, the first change is to 
stipulate that the function $A(r)$ provide a
dark shadow at multiple wavelengths.  Hence, the contrast
constraints are repeated for a discrete set of wavelengths that span the set
of wavelengths for which a shadow is
desired.  Of course then one needs to worry about the gaps between the chosen 
discrete set of wavelengths.  One possibility is simply to require the shadow
to be darker than necessary at the chosen wavelengths with the thought that
the performance can't degrade with arbitrary abruptness as one moves to
intermediate wavelengths.  But, a better solution is to impose smoothness
constraints on the function $A(r)$.  If this function is smooth, then one
expects the shadow to remain deep longer as one moves away from the specified
wavelengths.  A simple way to impose smoothness is to place a bound on
the magnitude of the second derivative
\[
    -\sigma \le A''(r) \le \sigma , \qquad 0 \le r \le R .
\]
Such constraints help, but
it turns out that the best thing to do is to let $\sigma$ be an optimization
variable and minimize this bound on the smoothness:
\begin{eqnarray}
    \mbox{minimize } & \sigma & \nonumber \\
    \mbox{subject to } 
      & -1 \le \Re(E_o(\rho)/(10^{-5}E_0/\sqrt{2})) \le 1,
        & \quad 0 \leq \rho \leq \rho_{\mathrm{max}} \nonumber \\
      & -1 \le \Im(E_o(\rho)/(10^{-5}E_0/\sqrt{2})) \le 1,
        & \quad 0 \leq \rho \leq \rho_{\mathrm{max}} \nonumber \\
      & -\sigma \le A''(r) \le \sigma , & \quad 0 \leq r \leq R \nonumber \\
      & 0 \leq A(r) \leq 1, & \quad 0 \leq r \leq R
    \label{op10}
\end{eqnarray}
(the original objective of minimizing the integral of $A(r)$ actually has
little effect on the problem as long as $R$ is small enough).
Of course, once we introduce a shadow constraint for each 
of several wavelengths, we
have the freedom to let the depth and width of the shadow be wavelength
dependent.

Practical considerations also provide further constraints.  For
a realistic binary occulter, the innermost section should be opaque
out to some radius $a$ to accommodate the spacecraft. This is
expressed as:
\begin{equation} \label{op13}
    A(r) = 1, \quad 0 \leq r \leq a
\end{equation}
We might also wish to impose the constraint that
$A'(r) \leq 0$ as this will ensure that the petal-mask to be described next
will have petals that get monotonically narrower as one moves out to the tip.
Such petal shapes are probably easier to manufacture.

\section{Adding petals} \label{sec:petal}

\begin{figure}
\begin{center}
\includegraphics[width=3in]{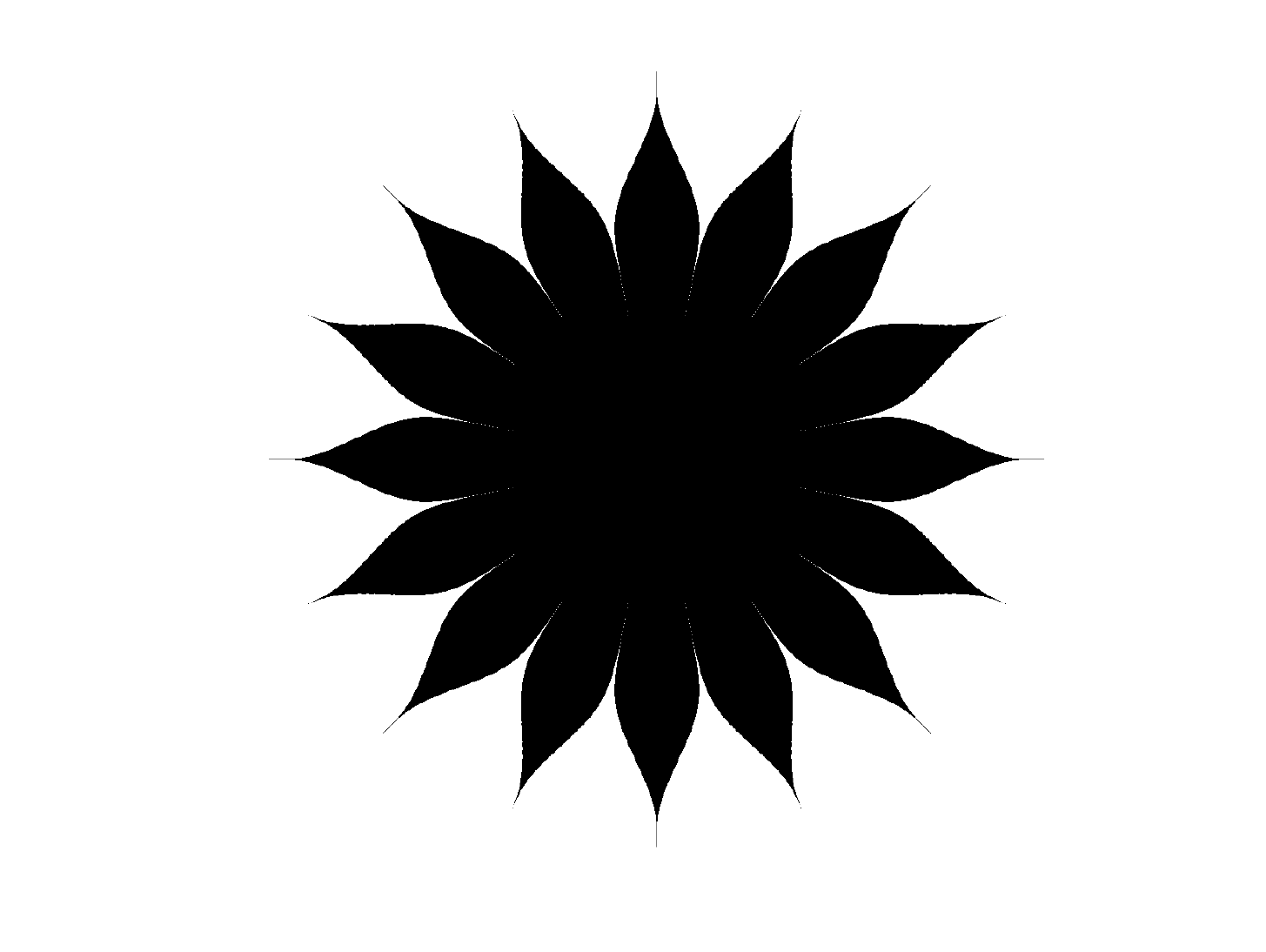}
\end{center}
\caption{An optimally-shaped sixteen-petal occulter.} \label{fig:1}
\end{figure}

Unfortunately, it is not currently possible to build an apodized occulter
to the required precision.  So, instead, we replace the apodized occulter
with a binary occulter of a particular shape.  
For instance, inspired by \citet{Van03}, \citet{Cas06} suggested using
an occulter made up of a
set of N identical evenly spaced ``petals'' as shown in Figure \ref{fig:1}.
These petals are
wedges of the circle whose width varies with radius such that the
fractional angular extent of the occulter at a given radius
is the attenuation profile $A(r)$.  
Except for Babinet's principle,
this petal-shaped occulter is identical to the starshaped pupil masks
described in \citet{Van03}.  The resulting propagated field for
such an occulter is thus found via the same procedure using the
Jacobi-Anger expansion.  The result is
\begin{eqnarray}
    E_{o\mathrm{,petal}}(\rho, \phi) &=& E_{o\mathrm{,apod}}(\rho) \nonumber \\
    &&- E_0 e^{\frac{2 \pi i z}{\lambda}}\sum^{\infty}_{j = 1}\frac{2 \pi (-1)^j}{i \lambda z}
    \left(\int^R_0 e^{\frac{\pi i}{\lambda
    z}(r^2+\rho^2)} J_{j N}\left(\frac{2 \pi r \rho}{\lambda
    z}\right) \frac{\sin{(j \pi A(r))}}{j \pi} r dr\right) \nonumber \\
    &&\qquad \qquad 
    \times \left(2\cos{(j N (\phi-\pi/2))}\right)\label{pet9}
\end{eqnarray}
where $E_{o\mathrm{,apod}}(\rho,z)$ is the field from the smooth
apodization and N is the number of petals (assumed even).  For large $N$, 
all of the $J_{jN}$ ($j > 0$) become small exponentially fast
near the center of the telescope and so the field approaches 
that of the smooth apodization as $N$ increases.

\section{Results} \label{sec:results}

One consideration that must be taken into account when designing
optimized occulters is angular size of the shade.  As mentioned in
Sec. \ref{sec:intro}, the maximum angular separation between
Earth-like planets and their stars is 0.1 arcsecond for a star 10
parsecs distant.  The angular size of the shade is $R/z$.  For
example, for a $25$ m radius shade, the shade must be at least
$51600$ km distant. If we want to see planets at smaller angular separations,
i.e., further from Earth, the shade must be shrunk or the distance
increased.  We present a series of shades optimized with different
sizes and at different distances.

Radially-symmetric apodizations were created to provide $10^{10}$
contrast out to a given radius for four separate occulter profiles:

$\cdot$ 18m occulter, 18000 km distance, 3m shadow radius

$\cdot$ 20m occulter, 36000 km distance, 3m shadow radius

$\cdot$ 25m occulter, 72000 km distance, 2.5m shadow radius

$\cdot$ 30m occulter, 100000 km distance, 2.5m shadow radius

\begin{figure}
\begin{center}
\includegraphics[width=3in]{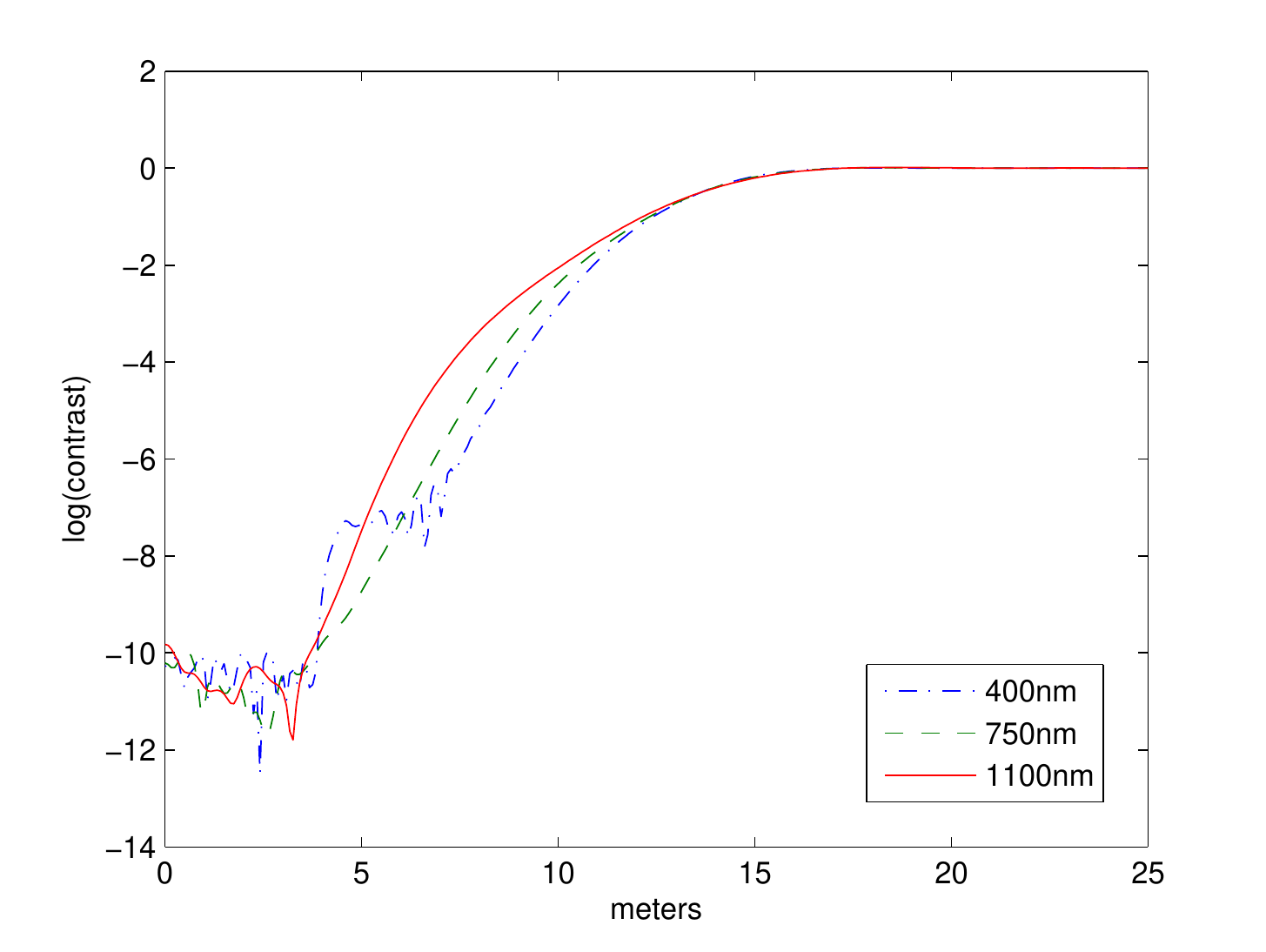}
\includegraphics[width=3in]{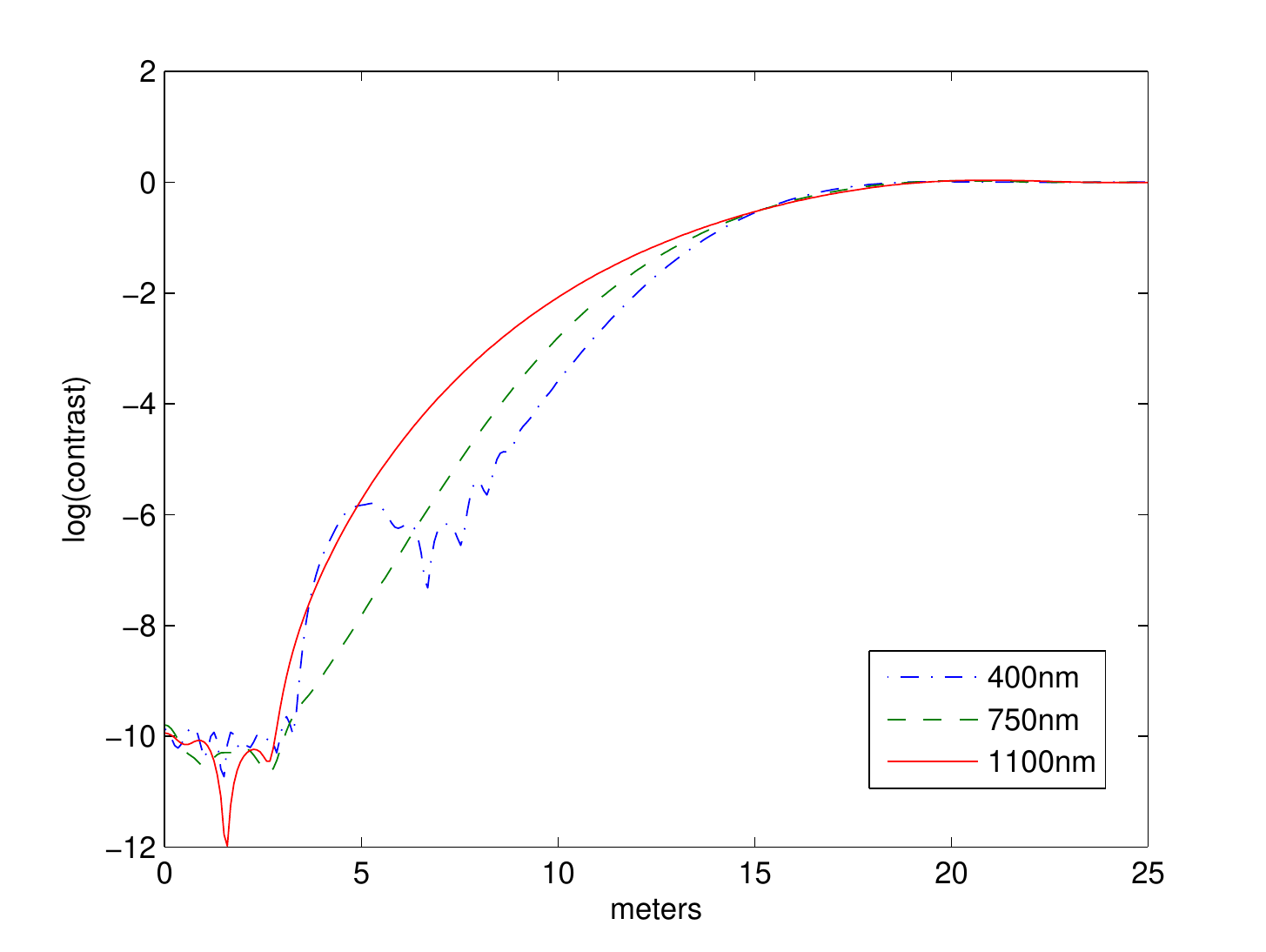}
\includegraphics[width=3in]{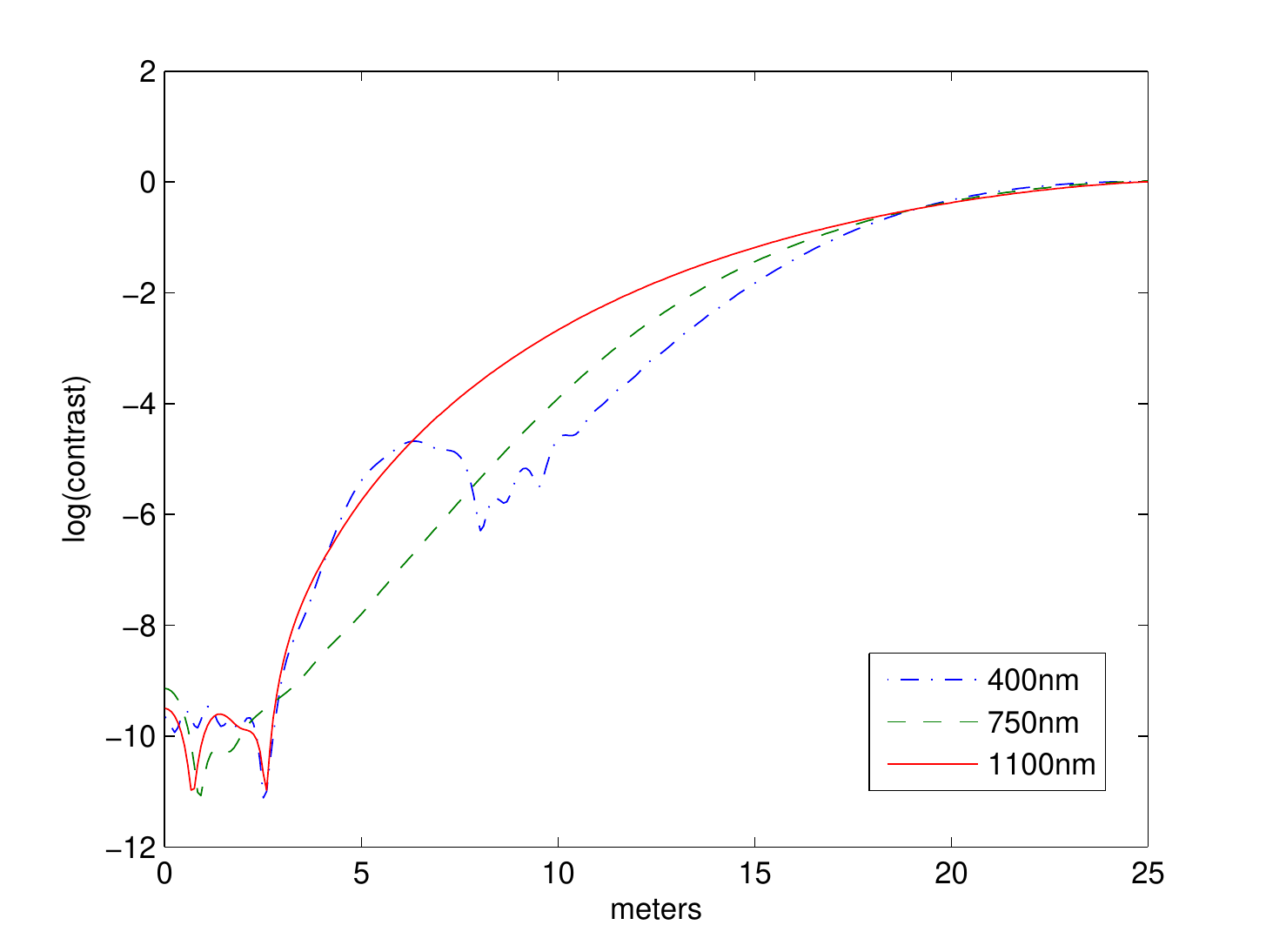}
\includegraphics[width=3in]{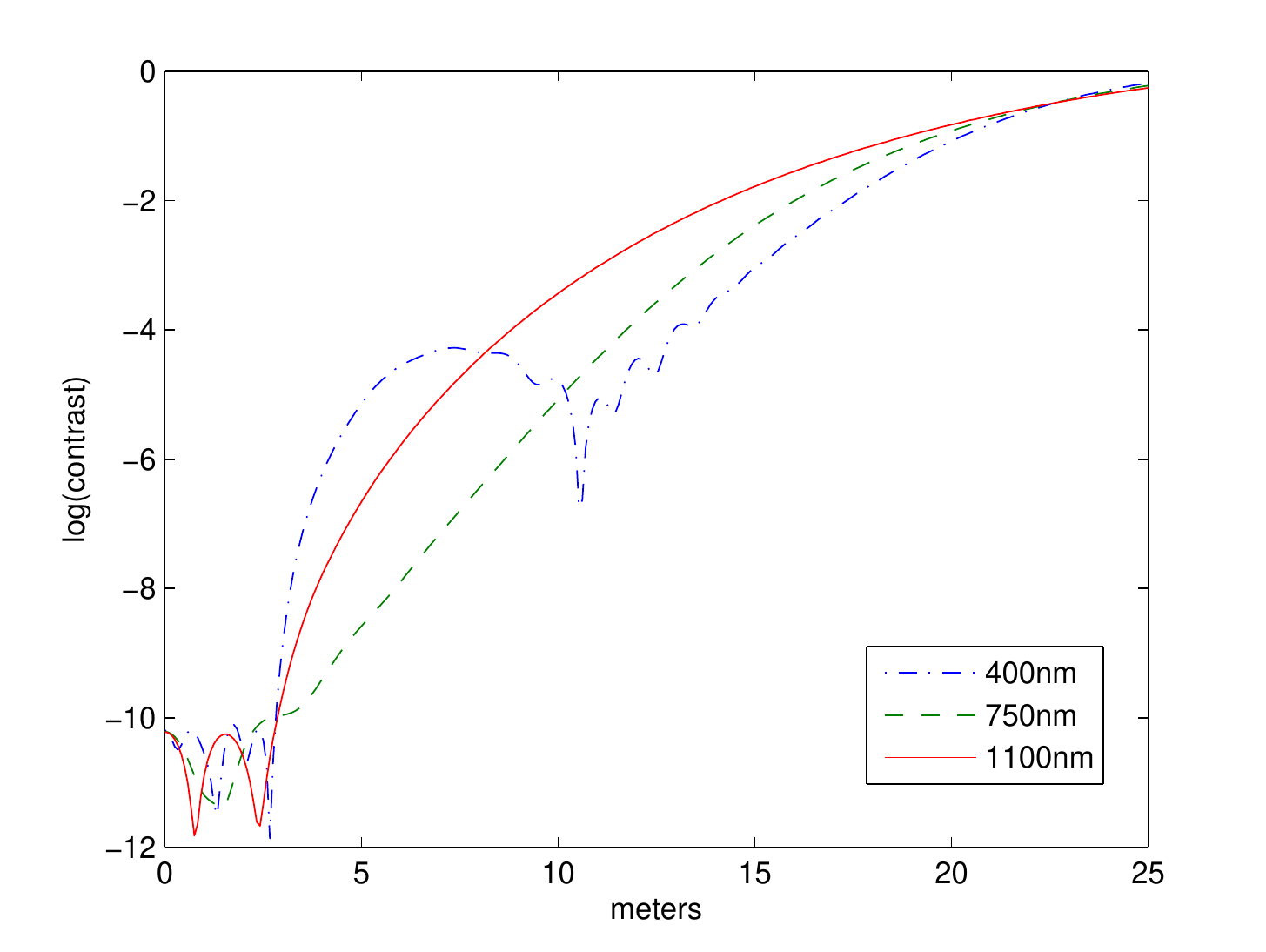}
\end{center}
 \caption{These plots show the radial profile of
  the shadow at the telescope, in different wavelengths.
  {\em Top left.} This plot is for an $18$m occulter at
  $18000$km.  {\em Top right.} This plot is for a $20$m
  occulter at $36000$km.
  {\em Bottom left.} This plot is for an $25$m occulter at
  $72000$km.  {\em Bottom right.} This plot is for a $30$m
  occulter at $100000$km.} \label{fig:2}
\end{figure}

These occulters were designed to provide the specified contrast over
a band from $400$nm to $1100$nm with contrast constraints specified in $100$nm
increments across this band.  Radial profiles of the shadow
at the telescope are shown in \myfig{fig:2}.  The profiles are shown for three
wavelengths: $400$nm, $750$nm, and $1100$nm.  Note that these wavelengths
correspond to the shortest and longest wavelengths at which high contrast was
dictated as well as an intermediate wavelength which happens to fall midway
between the two nearest wavelengths at which high contrast was constrained
($700$nm and $800$nm). 

\begin{figure}
\begin{center}
\includegraphics[width=1.6in]{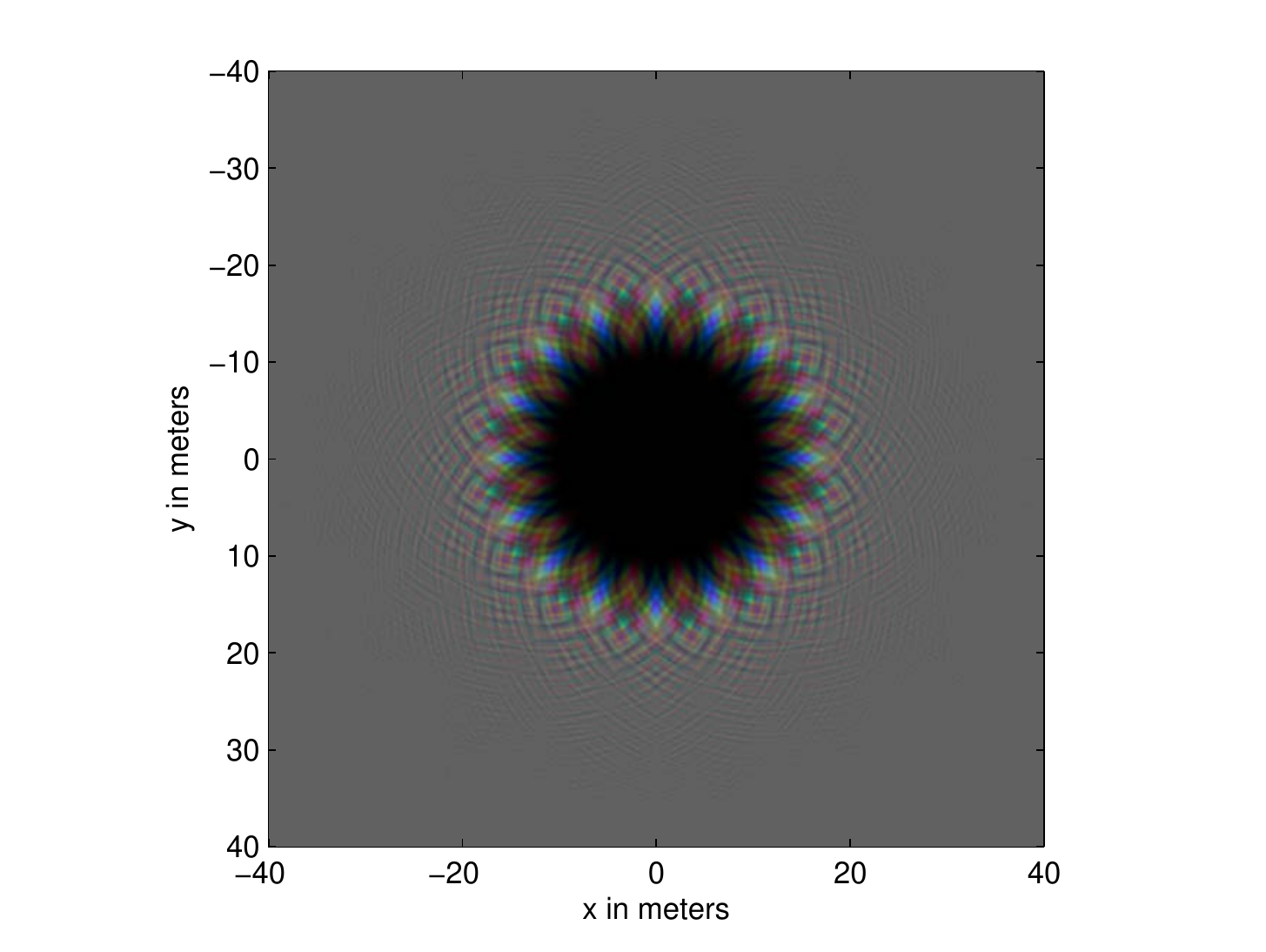} \hspace*{-0.2in}
\includegraphics[width=1.6in]{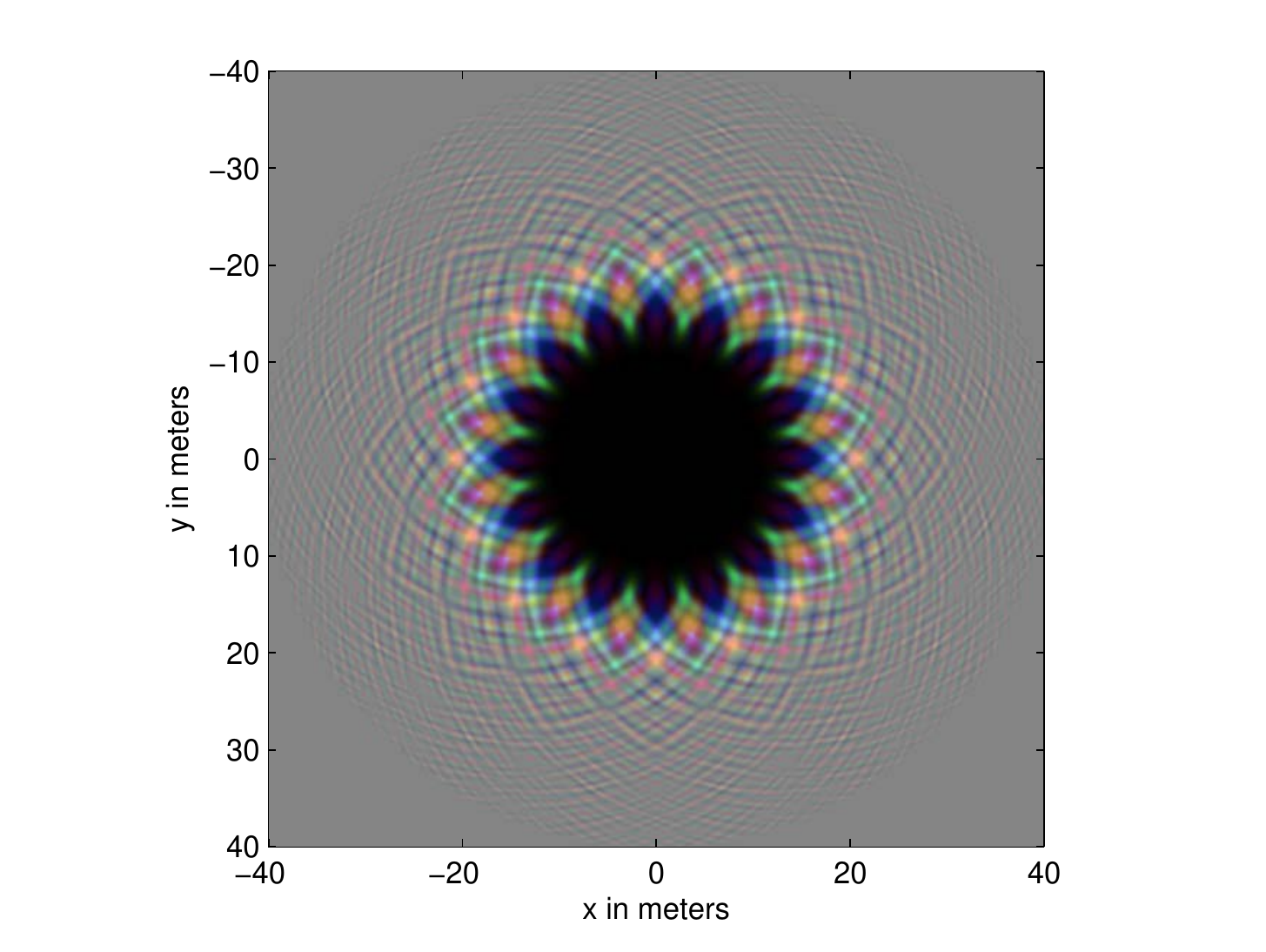} \hspace*{-0.2in}
\includegraphics[width=1.6in]{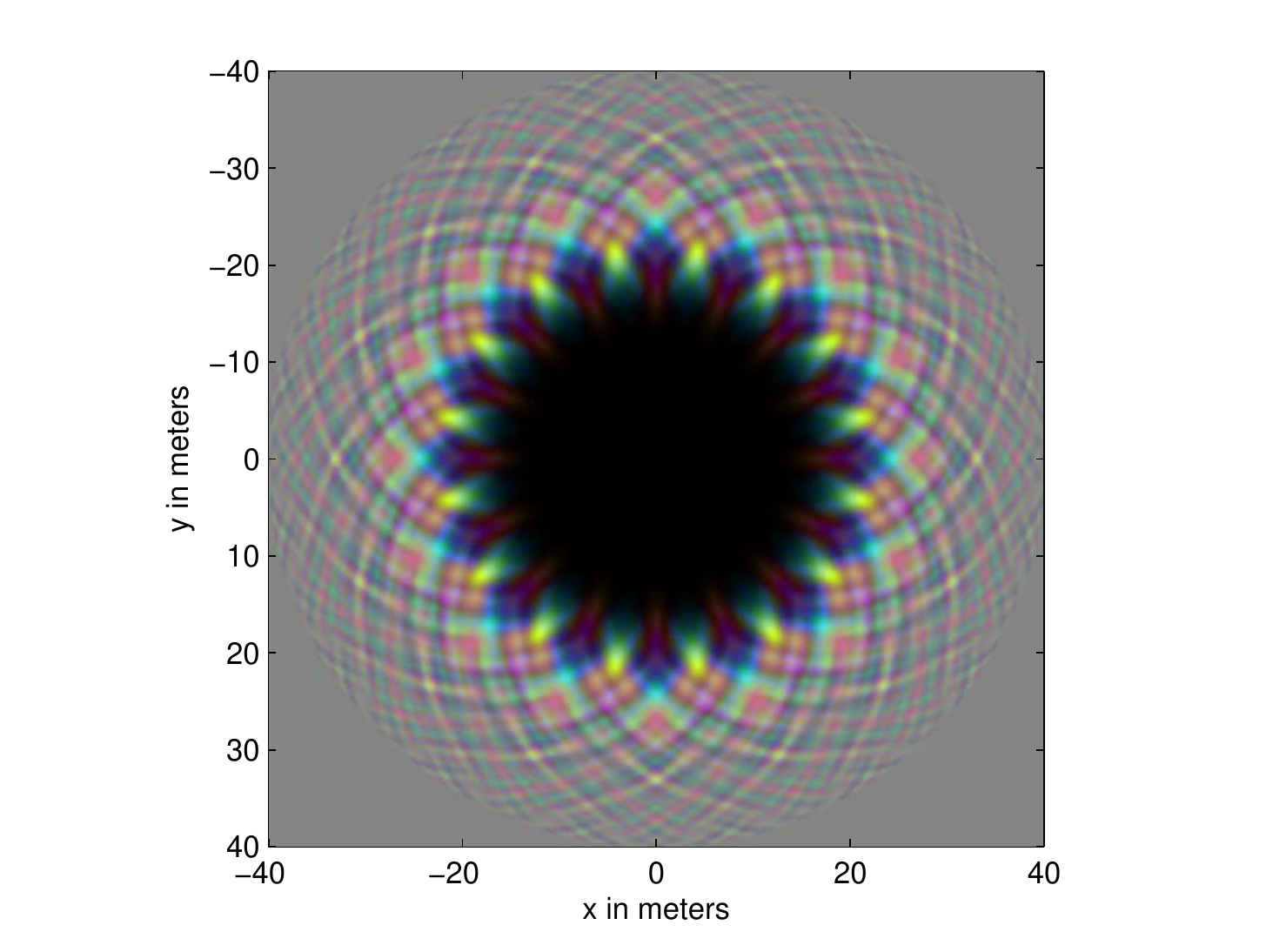} \hspace*{-0.2in}
\includegraphics[width=1.6in]{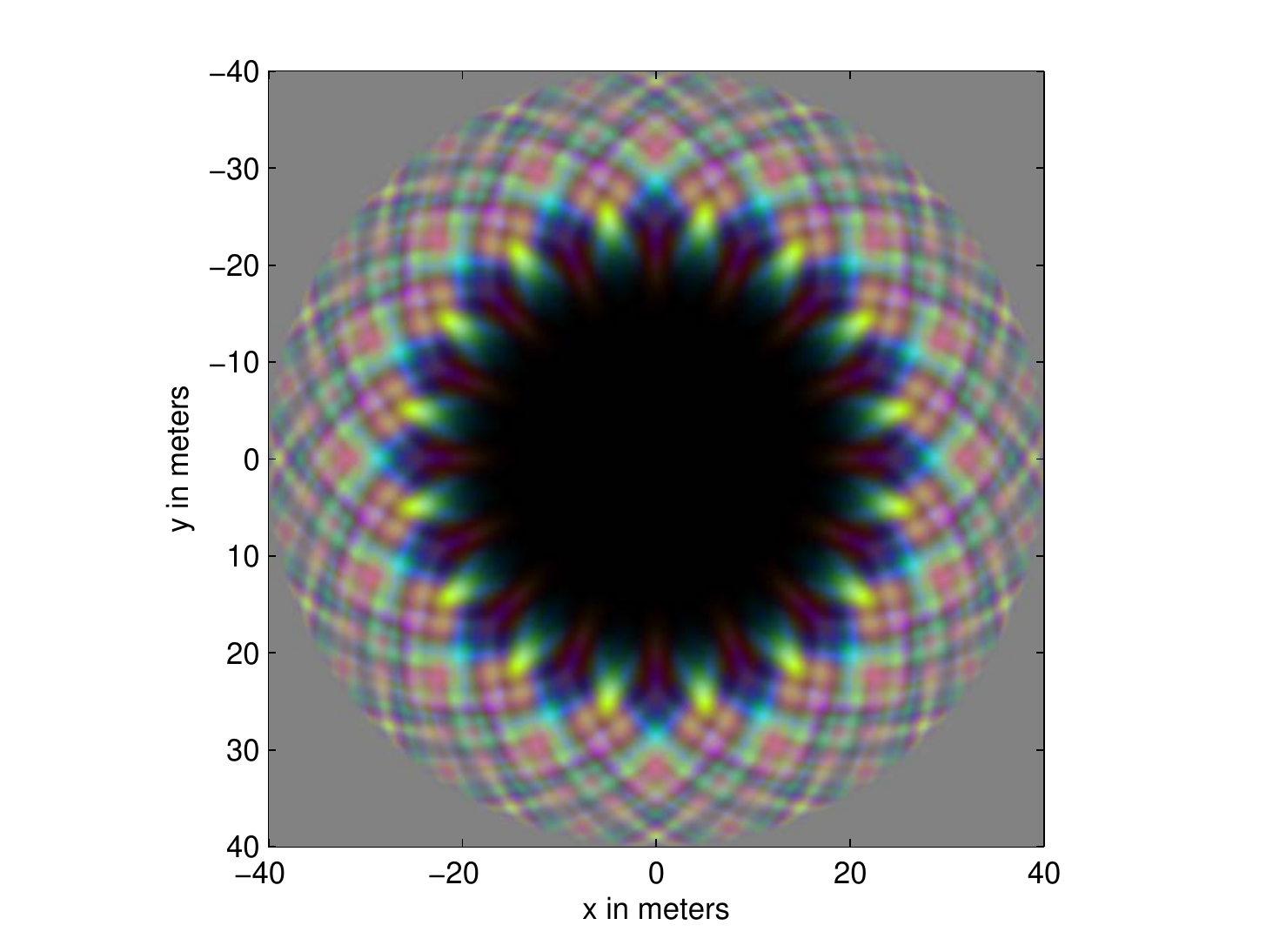} \\
\includegraphics[width=1.6in]{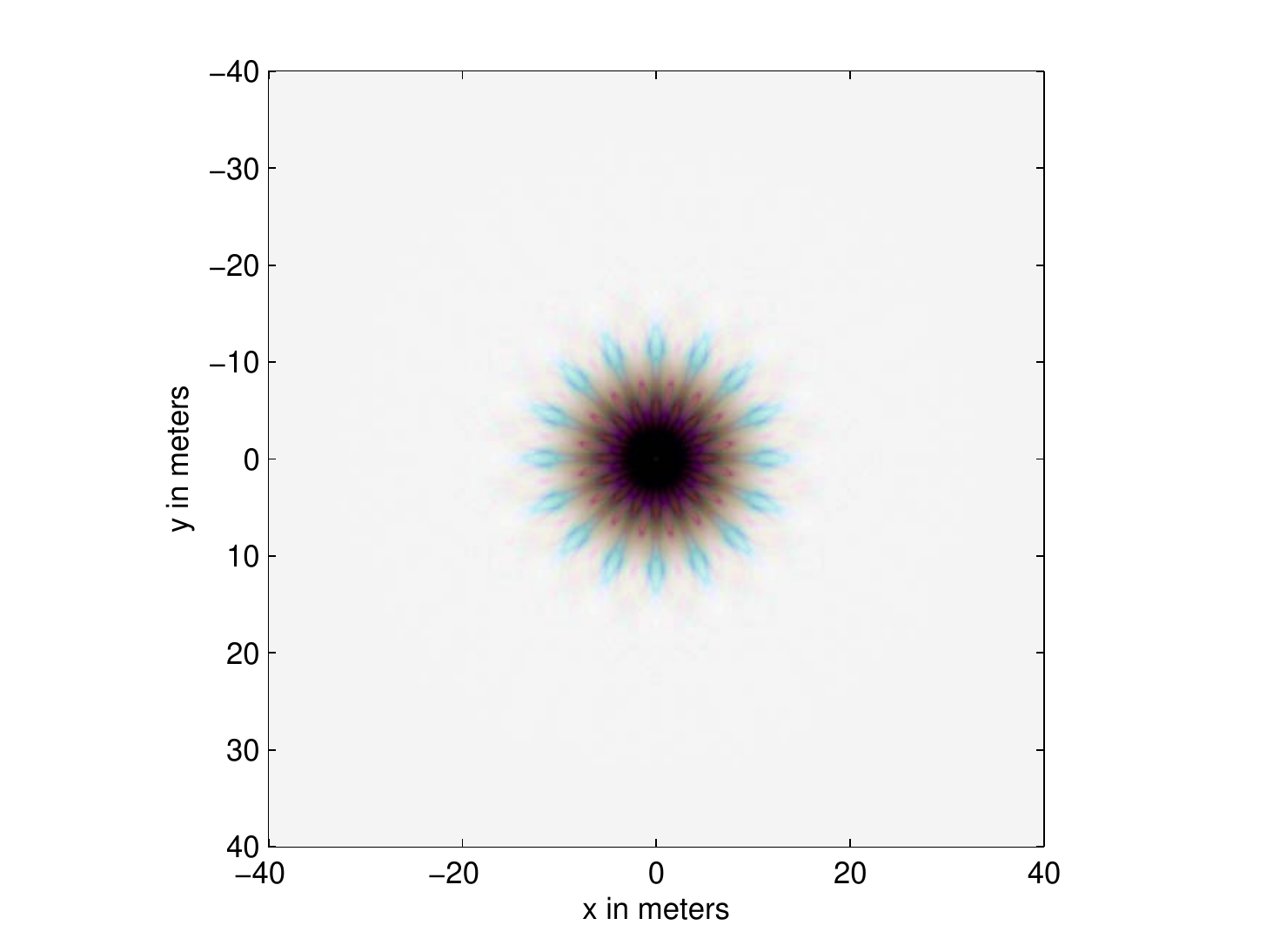} \hspace*{-0.2in}
\includegraphics[width=1.6in]{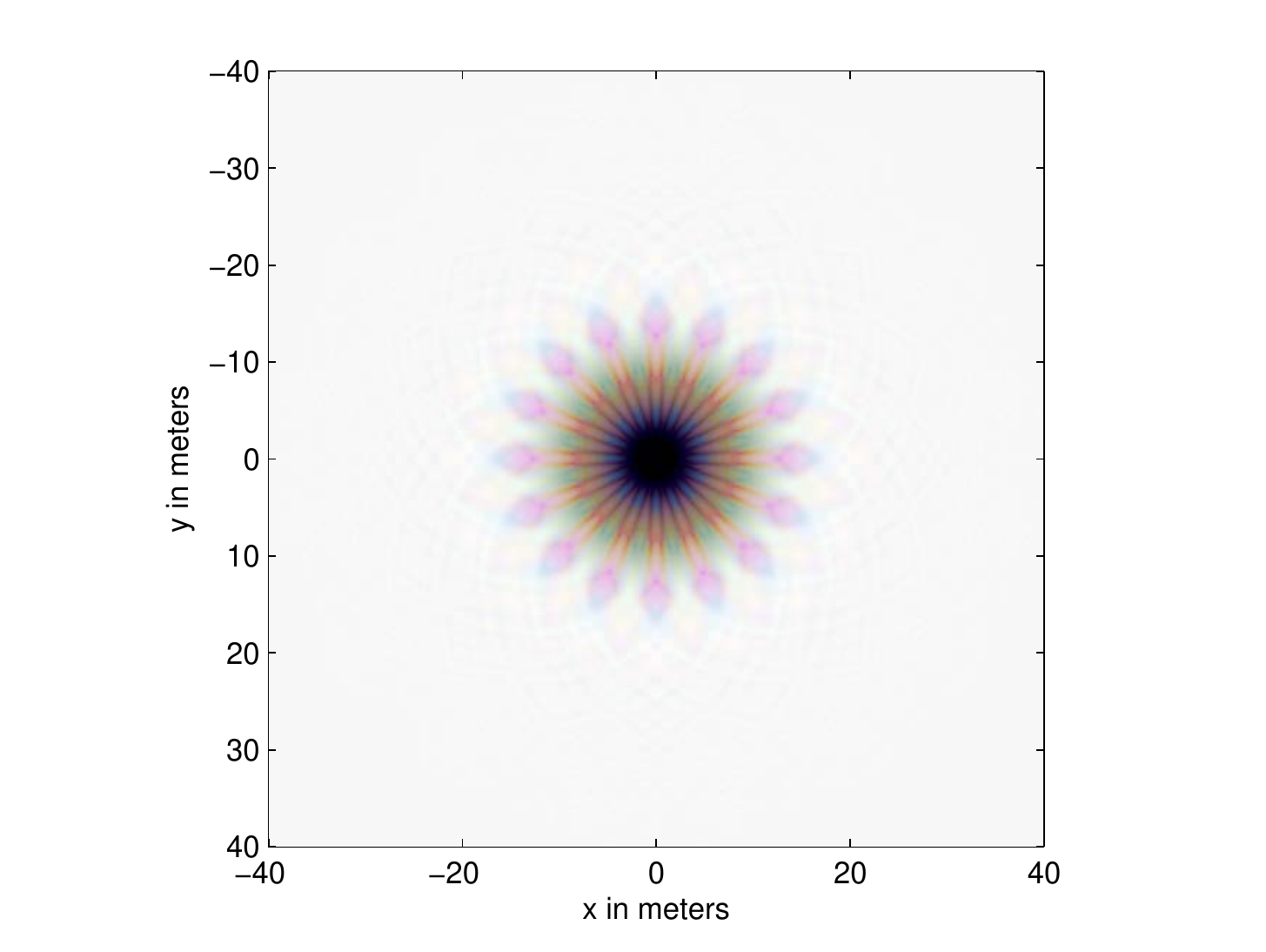} \hspace*{-0.2in}
\includegraphics[width=1.6in]{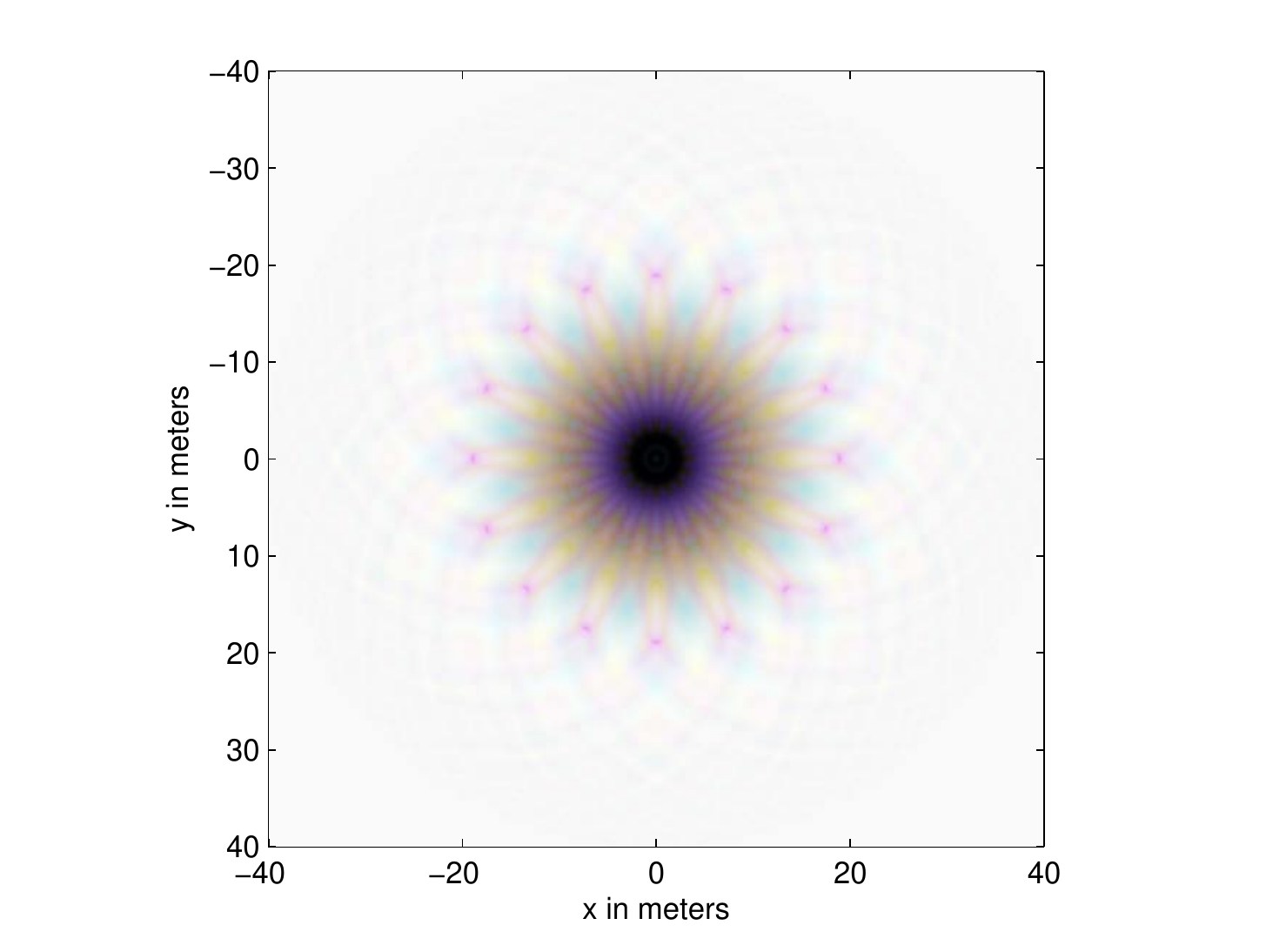} \hspace*{-0.2in}
\includegraphics[width=1.6in]{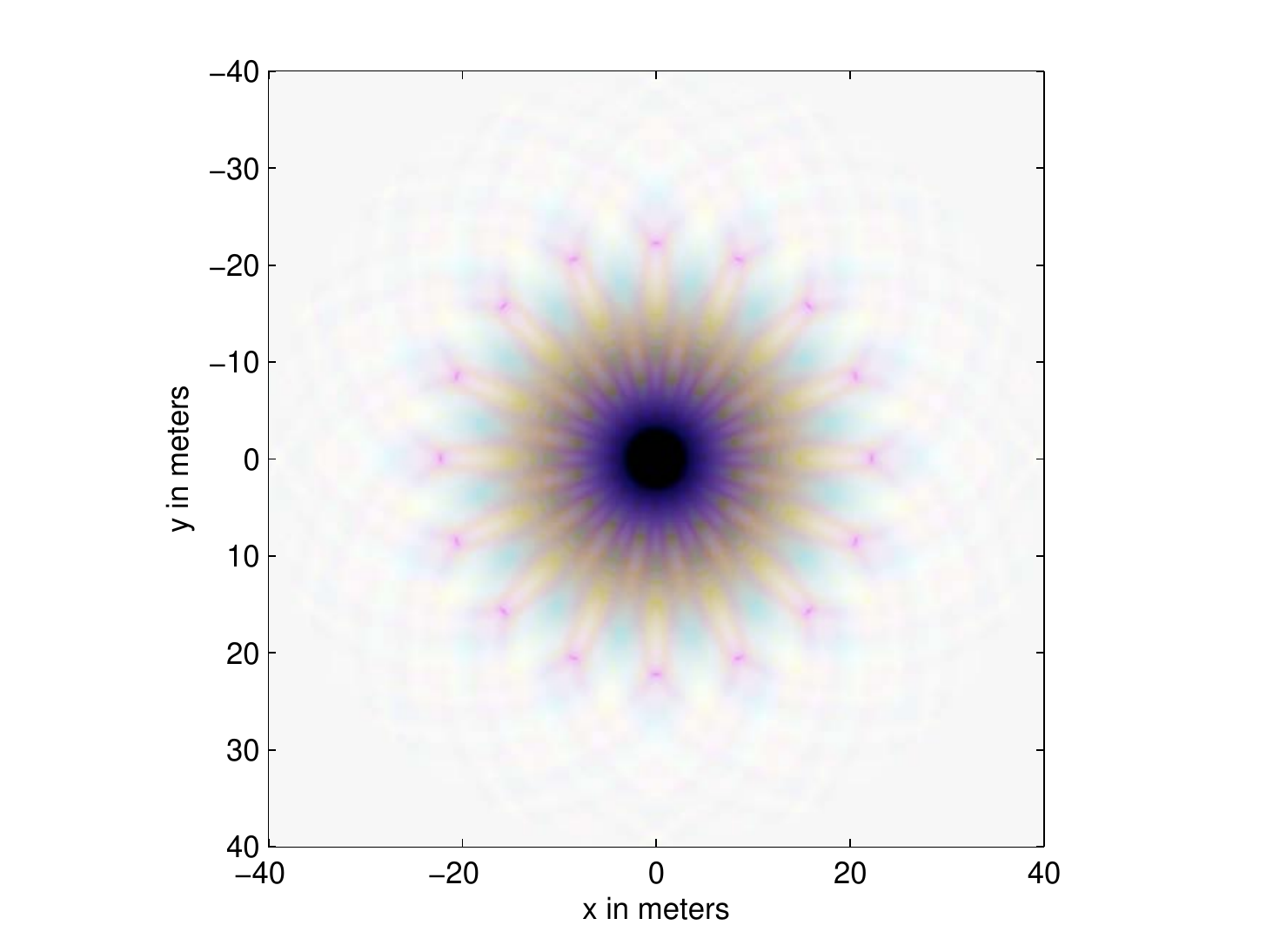}
\end{center}
 \caption{
 The shadow cast at the telescope pupil plane for four different occulter
 distances, which are from left to right
 $18,000$km, $36,000$km, $72,000$km, and $100,000$km.
 The top row shows linear stretch plots whereas the bottom row shows
 logarthmic stretches with $10^{-10}$ set to black.
 These are RGB images composited using $\lambda =
 1.0\mu$m for the red channel, $\lambda = 0.7\mu$m for the green
 channel, and $0.4\mu$m for the blue channel.  
  } \label{fig:3}
\end{figure}

Once a profile is created by optimization, we use \myeq{pet9} to
calculate the effect of converting a smooth apodization to petals;
this petalization tends to reduce the width of the shadow at certain
angles.  In a forthcoming paper, we will present a method of 
optimizing the petal shape directly, to prevent this degradation. Each
of the four occulters was converted to a binary occulter with $16$
petals; the performance of these occulters at $400$nm, $700$nm, and $1100$nm
is shown in \myfig{fig:3}.

\begin{figure}
\begin{center}
\includegraphics[width=3in]{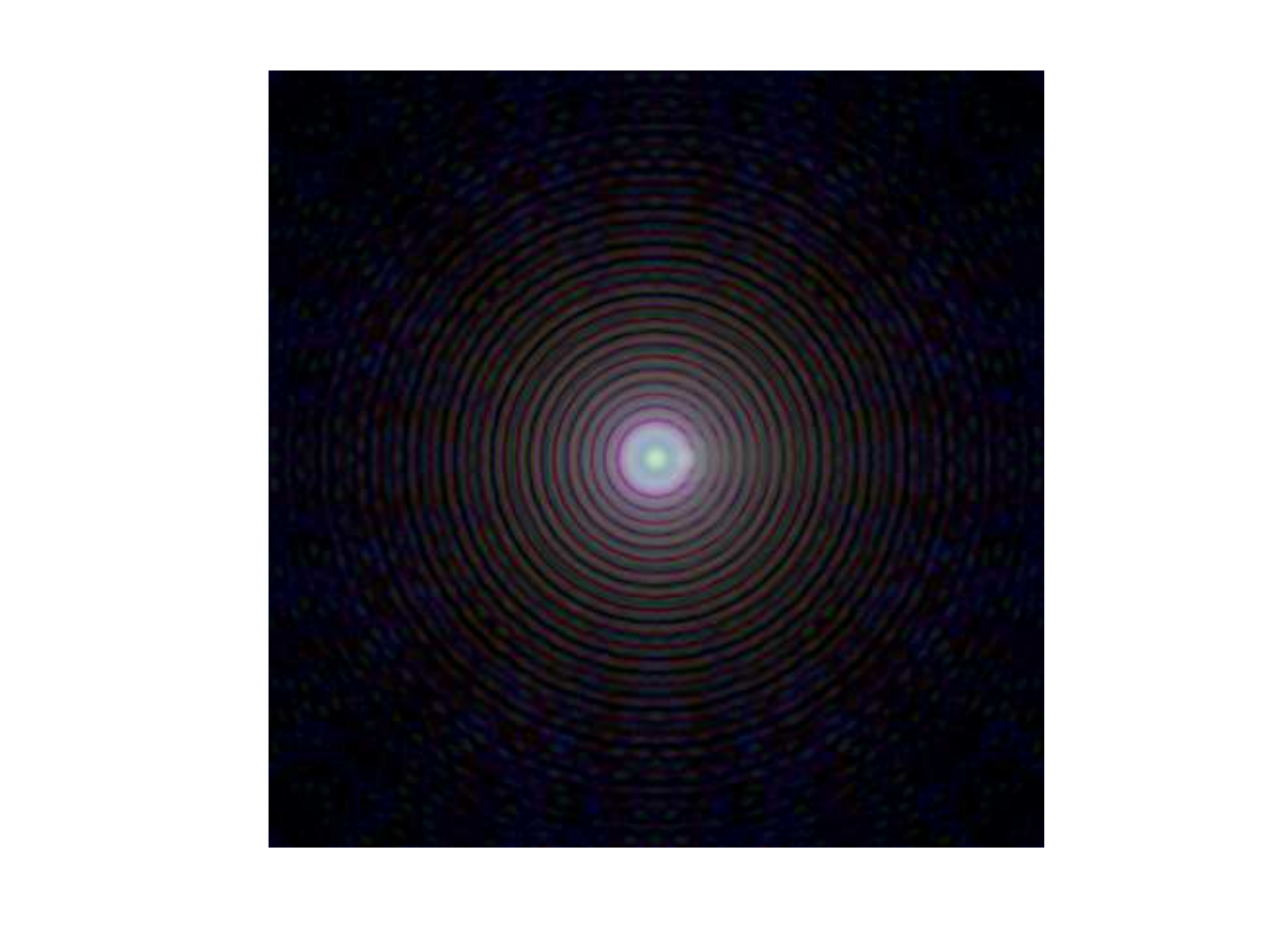}
\includegraphics[width=3in]{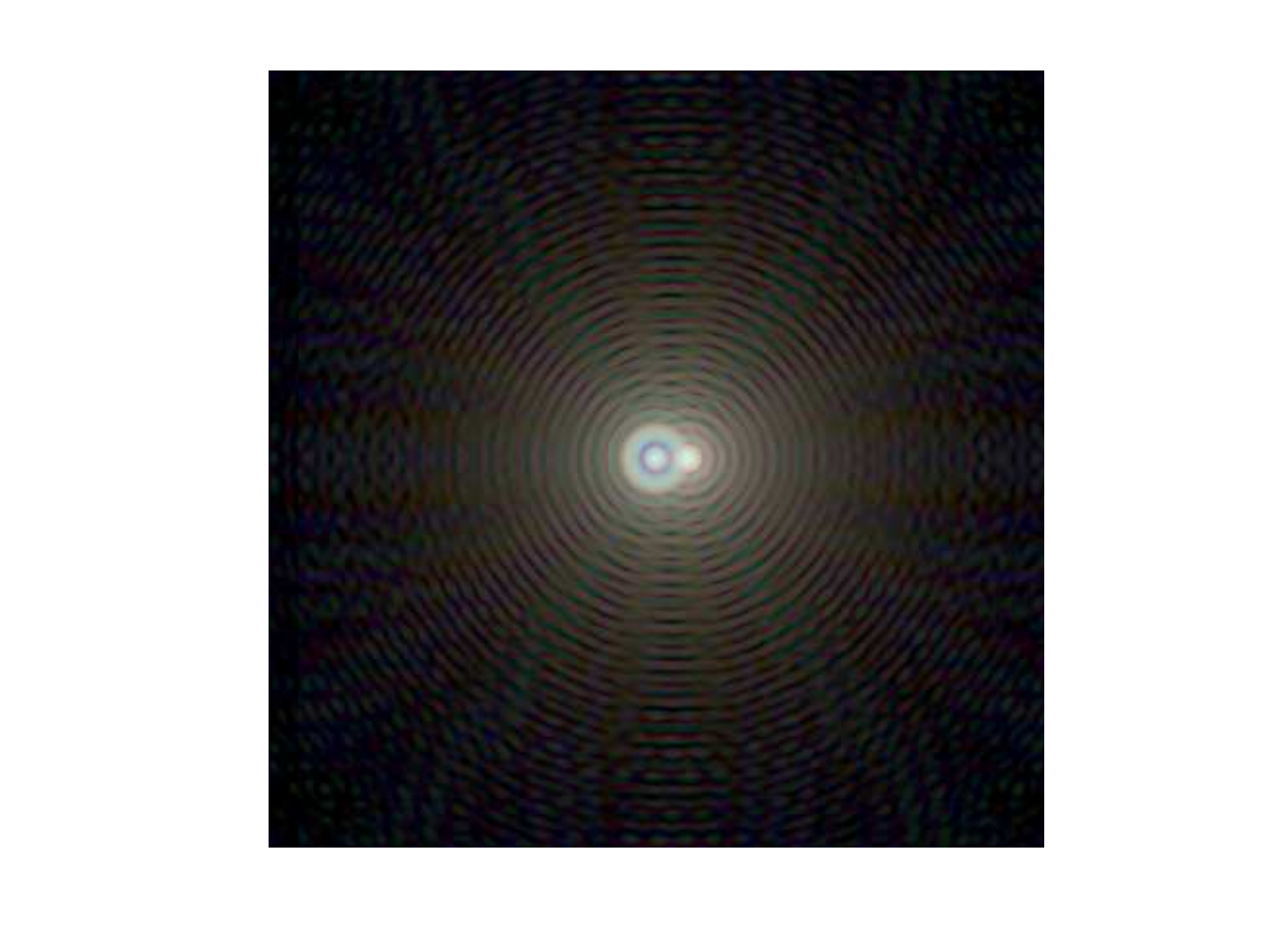}
\end{center}
 \caption{These plots are simulated (noiseless) images at 
 the telescope's image plane.  
 The RGB images were computed using $\lambda =
 1.0\mu$m for the red channel, $\lambda = 0.7\mu$m for the green
 channel, and $0.4\mu$m for the blue channel.  
 In both images, the off-axis planet is positioned at the tip of the the
 occulter.
  {\em Left.} This image is for an $R=22$m occulter at $66000$km.  
  The planet shown here is at $60$ milliarcseconds.
  {\em Right.} This plot is for an $R=25$m occulter at $72000$km.  
  The planet shown here is at $72$ milliarcseconds.
  } \label{fig:4}
\end{figure}

Finally, some may suggest that it is overly conservative to insist on
$10^{10}$ contrast at the telescope's pupil plane since additional contrast is
generated by the telescope itself as it forms an image.  The residual
starlight, being roughly flat across the telescope's pupil, forms something
similar to an Airy pattern in the image plane.  The planet will be slightly
off-axis and therefore offset slightly from the on-axis Airy pattern.  Since
the first diffraction ring in an Airy pattern is almost two orders of
magnitude suppressed relative to its main lobe, one can expect some benefit.
To test this, we modified our optimization code to
minimize an upper bound on the intensity of the light over a $6$m diameter 
shadow.  We ran tests assuming various separations $z$.  The tip radius $R$ 
was fixed so that the a planet appearing at the tip is $0.060$ arcseconds
off-axis (i.e., we set $R/z$ radians equal to $60$ milliarcseconds and solved 
for $R$).  The smallest value of $z$ that provides a sufficiently dark hole
for the planet to be detectable in the image plane turns out to be $66000$km.
For this case, the shadow at the telescope's pupil is slightly brighter than
$10^{-8}$ times the unattenuated brightness.  In the image plane, a planet
at $60$ milliarcseconds has about the same brightness as the residual 
starlight falling in the same location in the image (a $Q=1$ detection in TPF
parlance).  Figure \ref{fig:4} shows image plane images for the $66000$km
design described here.  Also shown in the figure for comparison is the image
plane image for the $72000$km design described earlier.

\section{Final Remarks} \label{sec:rem}

Whenever one uses optimization for engineering design, an important question
to address is this: how sensitive is the optimal design 
to small deviations from the given design scenario?
We have already discussed some of our efforts to ensure that our design is
robust.  Namely, we have discussed the issue of specifying shadow depth at
several wavelengths spread across the desired waveband and we have 
discussed using smoothness of $A(r)$ as a surrogate for solution robustness.
Furthermore, we have shown plots that verify the shadow depth at two
contrast-specified wavelengths ($400$nm and $1100$nm) as well as at a
wavelength at which contrast was not specifically constrained but instead is
midway between two such wavelengths.  In all three of these cases the depth of
the shadow proves to be more than adequate.

There are further robustness issues that need to be investigated.
For example, how deep will the dark shadow be if the occulter-telescope
separation deviates from the design value by a few percent?  Also, to what
precision do the petals need to be manufactured and then deployed?  Finally,
how much can the occulter's orientation be tilted relative to the
occulter-telescope axis?  
Regarding the second question,
preliminary analyses in which we randomly perturbed $A(r)$ by one part in
$100,000$ and recomputed the shadow profiles showed that perturbations at this
level do not degrade the depth or size of the shadow.  On the other hand,
perturbations at the level of one part in $10,000$ do start to affect
performance.  
Anyway, these are just very preliminary results.
All of the above questions are important and will be
addressed in detail in a forthcoming paper.  

In this paper we have used optimization techniques to investigate the
trade-off between inner working angle and size/distance of the occulter.
For terrestrial planet finding, it seems that the inner working angle should
be no more than 60 mas.  The number of Earth-like planets one can hope
to find drops quickly as one moves to larger separations.  We have shown that,
for an inner working angle 60 mas, the occulter needs to be about 50 m
tip-to-tip and it must be positioned about 72,000 km in front of the
telescope.  Future studies should be directed at determining 
whether such a size and
distance combination can be achieved within a reasonable mass and fuel budget.

%

\acknowledgments{The authors would like to thank R. Lyon for a
number of fruitful discussions.  We acknowledge support from the
Goddard Space Flight Center and Sigma Space Corporation, contract
\#NNG06EE69C.  R. Vanderbei acknowledges support from the ONR
(N00014-05-1-0206).}


\clearpage

\clearpage

\clearpage

\clearpage

\end{document}